\documentclass[pra,superscriptaddress,twocolumn,floatfix]{revtex4-1}
\usepackage{graphicx}
\usepackage{dcolumn}
\usepackage{bbm}
\usepackage{amsmath}
\usepackage{float}
\usepackage[english]{babel}
\usepackage[english]{babel}
\usepackage{dsfont}
\usepackage[normalem]{ulem}
\usepackage{xcolor}
 \usepackage{float}

\begin{document}
\title{Physics informed neural networks learning a two-qubit Hamiltonian}

\author{Leonardo K. Castelano}
\email{lkcastelano@.ufscar.br}
\affiliation{Departamento de F\'isica, Universidade Federal de S\~ao Carlos (UFSCar)\\ S\~ao Carlos, SP 13565-905, Brazil}

\author{Iann Cunha}
\affiliation{Departamento de F\'isica, Universidade Federal de S\~ao Carlos (UFSCar)\\ S\~ao Carlos, SP 13565-905, Brazil}
\author{Fabricio S. Luiz}
\affiliation{Faculty of Sciences, UNESP - S{\~a}o Paulo State University, 17033-360 Bauru, S{\~a}o Paulo, Brazil}
\affiliation{Federal Institute of São Paulo, 18202-000 Itapetininga, S{\~a}o Paulo, Brazil}
\author{Marcelo V. de Souza Prado}
\affiliation{Faculty of Sciences, UNESP - S{\~a}o Paulo State University, 17033-360 Bauru, S{\~a}o Paulo, Brazil}
\author{Felipe F. Fanchini}
\affiliation{Faculty of Sciences, UNESP - S{\~a}o Paulo State University, 17033-360 Bauru, S{\~a}o Paulo, Brazil}

\date{\today}
\begin{abstract}
Machine learning techniques are employed to perform the full characterization of a quantum system. The particular artificial intelligence technique used to learn the Hamiltonian is called physics informed neural network (PINN). The idea behind PINN is the universal approximation theorem, which claims that any function can be approximate by a neural network if it contains enough complexity. Consequently, a neural network can be a solution of a physical model. Moreover, by means of extra data provided by the user, intrinsic physical parameters can be extracted from the approach called inverse-PINN. Here, we apply inverse-PINN with the goal of extracting all the physical parameters that constitutes a two qubit Hamiltonian. We find that this approach is very efficient. To probe the robustness of the inverse-PINN to learn the Hamiltonian of a two-qubit system, we use the IBM quantum computers as experimental platforms to obtain the data that is plugged in the PINN. We found that our method is able to predict the two-qubit parameters with 5\% of accuracy on average. 
\end{abstract}
\maketitle

\section{Introduction}
A significant amount of effort has been invested in the development of new technologies aimed at advancing the field of quantum computing due to its potential to revolutionize digital computing. Precise knowledge of the quantum system, described by a Hamiltonian containing both local and non-local terms, is essential for performing quantum operations and subsequently implementing quantum algorithms. Therefore, the extraction of this information from experimental data becomes a crucial task.
Interest in the application of artificial intelligence (AI) to address scientific challenges has been rapidly growing~\cite{Alipanahi2015, LeCun2015, doi:10.1126/science.aab3050, NIPS2012_c399862d}.The use of AI in physics began with the analysis of particle physics experiments~\cite{DENBY1988429,Barateetal.1999,Radovic2018}. Other applications include the application of machine learning (ML) in condensed matter systems \cite{Bedolla_2021,Juan_Carrasquilla}, exploring the AdS/CFT correspondence \cite{PhysRevD.98.046019}, and phase transition determination \cite{PhysRevB.100.045129}.

More recently, the concept of physics-informed neural network (PINN) has been introduced, where differential equations describing the physics of the problem are introduced in the training of the neural network~\cite{RAISSI2019686,doi:10.1126/sciadv.abi8605,doi:10.1137/19M1274067}. This approach offers the advantage of reducing the amount of training data required, as the neural network is constrained to satisfy the differential equations that adhere to physical laws ~\cite{RAISSI2019686,doi:10.1126/sciadv.abi8605,doi:10.1137/19M1274067}. Furthermore, the idea of learning and extracting information from a collection of data can also be implemented through the inverse-PINN~\cite{RAISSI2019686,doi:10.1126/sciadv.abi8605,doi:10.1137/19M1274067}. In this case, data is provided along with the equations and physical parameters can be extracts from the model.

Several approaches to learn a Hamiltonian have been proposed in the past. Some rely on the tomography of the density matrix ~\cite{doi:10.1080/09500349708231894,PhysRevLett.90.193601,PhysRevA.87.062119}, while other protocols focus on utilizing states that inherently encode information about the Hamiltonian, such as steady states and thermal states ~\cite{Anshu2021,PhysRevLett.122.020504}.Various learning proposals have been put forward that eliminate the need for preparing specialized initial states ~\cite{PhysRevLett.124.160502}. For instance, the parameters of the Hamiltonian can be derived from the Ehrenfest theorem~\cite{zubida2021optimal} or from measuring state properties through their evolution to identify the nearest-neighbor coupled Hamiltonian in superconducting systems ~\cite{hangleiter2021precise}. Yu at al.~\cite{Yu2023robustefficient} proposed a method that utilizes short-time Hamiltonian evolution and exploits ideas from randomized
benchmarking~\cite{10.1145/3408039}.

In this paper, we apply the technique of inverse-PINN to perform the Hamiltonian tomography for a two-qubit system. Experimental data at specific points is also required to extract these parameters. As is well-known, measurements in quantum mechanics are probabilistic and necessitate repeated preparation of the initial configuration for statistical measurement. If measurements are performed as a function of time, statistical measurements must be conducted for each time step. Therefore, probing all observables as a function of time requires a significant number of repetitions of the experiment. To address this issue, we investigate the accuracy of inverse-PINN as a function of the number of collocation points for density matrix tomography measurements. Here, collocation points refer to the points in the time domain where density matrix tomography is performed. We found that inverse-PINN can accurately provide the coupling between the qubits with a reduced number of collocation points. We also incorporate errors into our theoretical model for density matrix tomography to estimate the robustness of the predicted parameters.
Finally, we implemented an experiment on IBM quantum computers to address a real-world problem, and we successfully learned the two-qubit Hamiltonian with an error of less than 5\% on average, using only 20 collocation data points.

\section{Theoretical Model}
The general Hamiltonian for two-qubit can be written as
\begin{equation}
   H = -\frac{\hbar}{2} \sum^3_{l=0}\sum^3_{k=0}J_{k,l} \sigma_{k} \otimes  \sigma_{l},
\label{eq:hamiltiniano_ML}
\end{equation}
where $\sigma_{k}$ denote the corresponding Pauli matrix for $k = 1, 2, 3$, and $\sigma_{0}$ is the identity matrix. There are fifteen $J_{k,l}$ terms that describe local and non-local interactions between the two-qubit. The term $J_{0,0}$ only provides a reference for the energy and we set it equal to zero. The expected value for the corresponding observable can be calculated from
\begin{equation}
    \langle \sigma_{k}\sigma_{l} \rangle(t) = \mathrm{Tr}[\rho(t) \sigma_{k}\otimes  \sigma_{l}],
    \label{eq:measurement_trace}
\end{equation}
where $k,l \in\{0,3\}$ and $\rho(t)$ is the density matrix for the two-qubit at time $t$. Moreover, the dynamics of the observables must be included in the inverse-PINN, thus we use the Heisenberg equation
\begin{equation}
    \frac{d\langle \hat{O}_m \rangle(t)}{dt}= \frac{i}{\hbar} \langle[H,\hat{O}_m]\rangle.
    \label{eq:schrodinger_dynamics}
\end{equation}
The above equation for $\langle\hat{O}_m\rangle$, must be implemented for all 15 physical terms corresponding to all $\sigma_{k}\sigma_{l}$ different from identity. In this case, we are let to 15 coupled ordinary differential equations that must be solved for an initial condition to obtain the desired parameters.

The idea behind PINN is to model the solutions of the differential equations by a neural network and to minimize the loss function. Particularly, the loss function can be written as
\begin{equation}
    L=L_{model}+ L_{data},
\end{equation}
where
\begin{equation}
    L_{model}=\sum_{m=1}^{15}\sum_{j=1}\left|\left.\left(\frac{d\langle \hat{O}_m \rangle}{dt}- \frac{i}{\hbar} \langle[H,\hat{O}_m]\rangle\right)\right|_{t_j}\right|^2.
    \label{eq:Lmodel}
\end{equation}
The loss function associated to the model $L_{model}$ provides the values of $\langle\hat{O}_m\rangle$ at the collocation points $t_j$, which are mapped onto the NN, consequently the solutions of the differential equations are represented by the NN. The loss function $L_{data}$ is related to the data extracted from the experiment. This loss function impose the constraint for the solutions of the differential equations fit the experimental data in the collocation points. In this sense, the inverse-PINN forces the NN to be both the solutions of the differential equations and to represent the experimental data, thereby performing the Hamiltonian tomography and extracting the physical parameters $J_{k,l}$. 

\section{Results}
We start by analyzing  two simpler models, the $H_Z$ and the $H_{XYZ}$ Hamiltonians. The first Hamiltonian is
\begin{equation}
    H_Z=-\frac{\hbar}{2}\left(J_{0,3}\sigma_0\otimes\sigma_3+J_{3,0}\sigma_3\otimes\sigma_0+J_{3,3}\sigma_3\otimes\sigma_3\right),\label{eq:HZ}
\end{equation}
which has been used to fit experimental data related to quantum dots \cite{shulman2012demonstration}. The second Hamiltonian is the XYZ model without local terms, thus
\begin{equation}
    H_{XYZ}=-\frac{\hbar}{2}\left(J_{1,1}\sigma_1\otimes\sigma_1+J_{2,2}\sigma_2\otimes\sigma_2+J_{3,3}\sigma_3\otimes\sigma_3\right).\label{eq:HXYZ}
\end{equation}
\begin{figure}[b]
    \centering
    \includegraphics[width=0.5\textwidth]{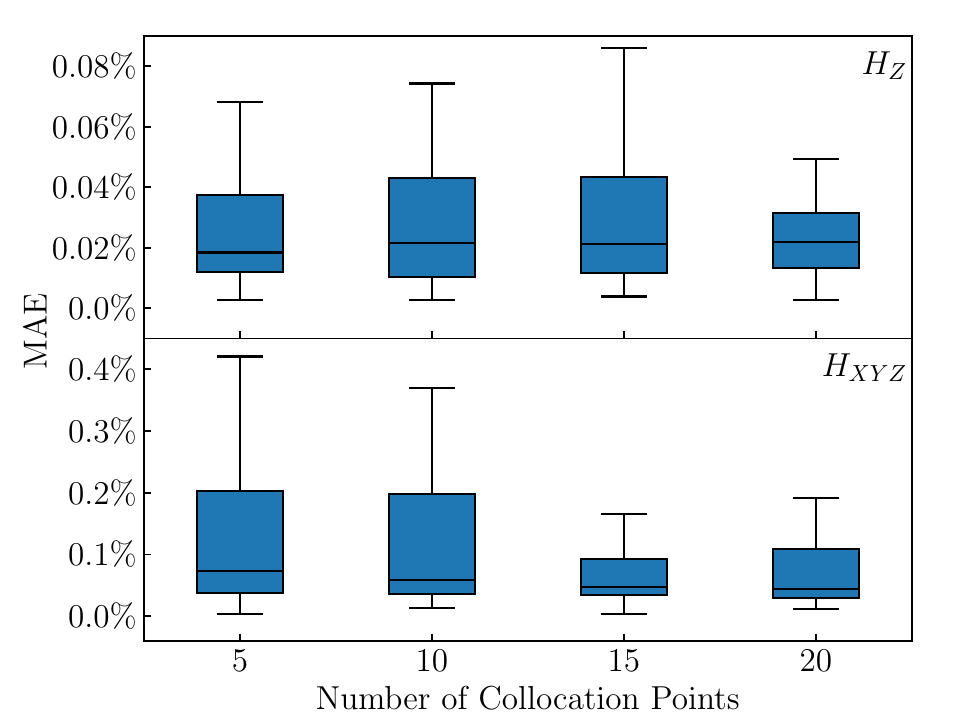}\label{fig1}
    \caption{MAE for the $H_Z$ Hamiltonian (top panel) and for the $H_{XYZ}$ Hamiltonian (bottom panel) as a function of the number of collocation points.}
    \end{figure}
Both cases have the advantage of decoupling some differential equations, thus there are only four coupled ode's for $J_{0,3}$ and $J_{3,3}$ ($J_{1,1}$ and $J_{3,3}$). Symmetric equations for $J_{3,0}$ and $J_{3,3}$ ($J_{2,2}$ and $J_{3,3}$) complete the set of ode's. Because these equations are symmetric, we can only solve a set of four ode's to study the results for these cases.
First, we make an analysis concerned on the number of points of experimental data needed. To perform such a task, we provide data without errors by numerically solving equations~\ref{eq:schrodinger_dynamics} considering the values of the parameters $J_{k,l}$ randomly sorted between $[-\omega_0,\omega_0]$, where $\omega_0=2\pi/T$ and $T$ is the final time of evolution. The first analysis that we perform concerns on accuracy of the extracted parameter versus the number of collocations points. The accuracy is measured by the mean absolute error, which is defined as follows:
\begin{equation}
    MAE=\frac{1}{D}\sum_{i=1}^D\frac{|P^{exact}_i-P^{pred}_i|}{|P^{exact}_i|},
\end{equation}
where $D$ is the number of physical parameters, $P^{exact}_i$  ($P^{pred}_i$)  denotes the i-th exact (predicted) physical parameter . In Figure \ref{fig1}, we plot the MAE as a function of the number of collocations points in time for both Hamiltonians $H_Z$ (top panel) and $H_{XYZ}$ (bottom panel). We use $D=50$ in both cases and we plot the results using the boxplot method to show the spread of the data. Thus,  the lowest (highest) marker shows the lowest (highest) data point in the data set excluding any outliers, which are data points that differ significantly from other observations. The lowest (highest) marker of the blue box is related to the lower (higher) quartile and the marker inside the blue box indicates the median. Figure \ref{fig1} demonstrates that MAE is lower than 0.4\% for only 5 collocation points, which is a evidence for performing the Hamiltonian tomography with very good accuracy.

To further analyze the performance of the Hamiltonian tomography, we add a random Gaussian error in the parameters, which is characterized by the standard deviation $\sigma$. In this case, parameters $J_{k,l}$ in equation~(\ref{eq:hamiltiniano_ML}) are modified according to $J_{k,l}\rightarrow J^0_{k,l}+E_{k,l}$, where $E_{k,l}$ is a random variable with Gaussian distribution characterized by $\sigma$. The idea is to generate the input data for the inverse-PINN considering the error $E_{k,l}$ and try to predict the value $J^0_{k,l}$ without error.
\begin{figure}[t]
    \centering
    \includegraphics[width=0.5\textwidth]{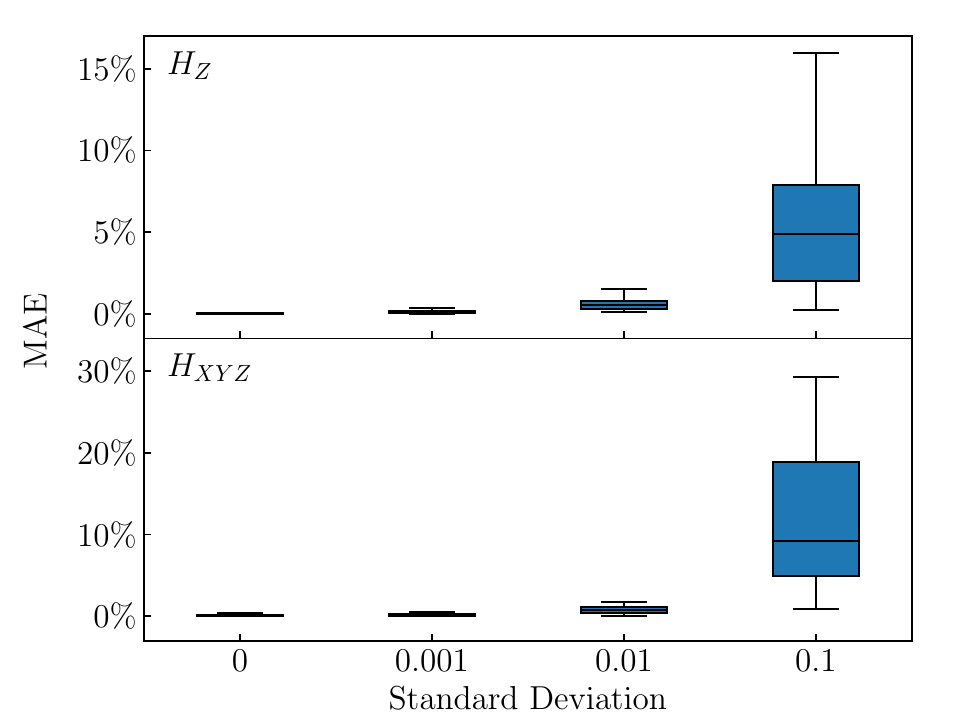}
    \caption{MAE for the $H_Z$ Hamiltonian (top panel) and for the $H_{XYZ}$ Hamiltonian (bottom panel) as a function of the standard deviation for N=5 collocation points.}
    \label{fig2}
\end{figure}

\begin{figure}[b]
    \centering
    \includegraphics[width=0.5\textwidth]{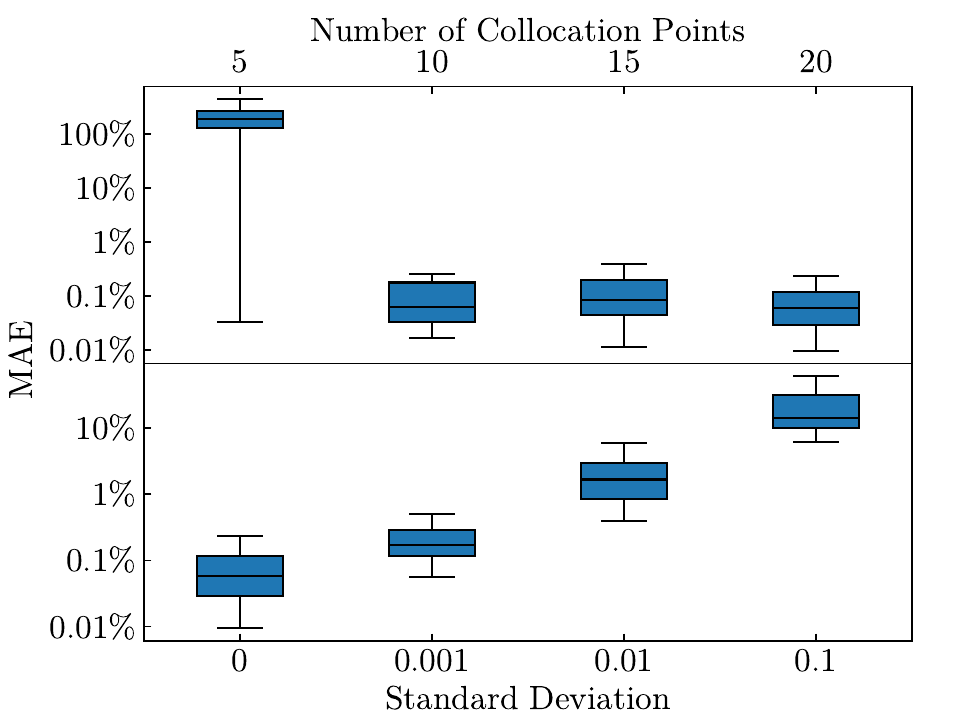}
    \caption{MAE as a function of the number of collocation points (top panel). MAE  as a function of the standard deviation for N=20 collocation points (bottom panel). Both panels are related to the general two qubit Hamiltonian and are plotted in logarithm scale. }
    \label{fig3}
\end{figure}

\begin{figure}[t]
    \centering
    \includegraphics[width=0.5\textwidth]{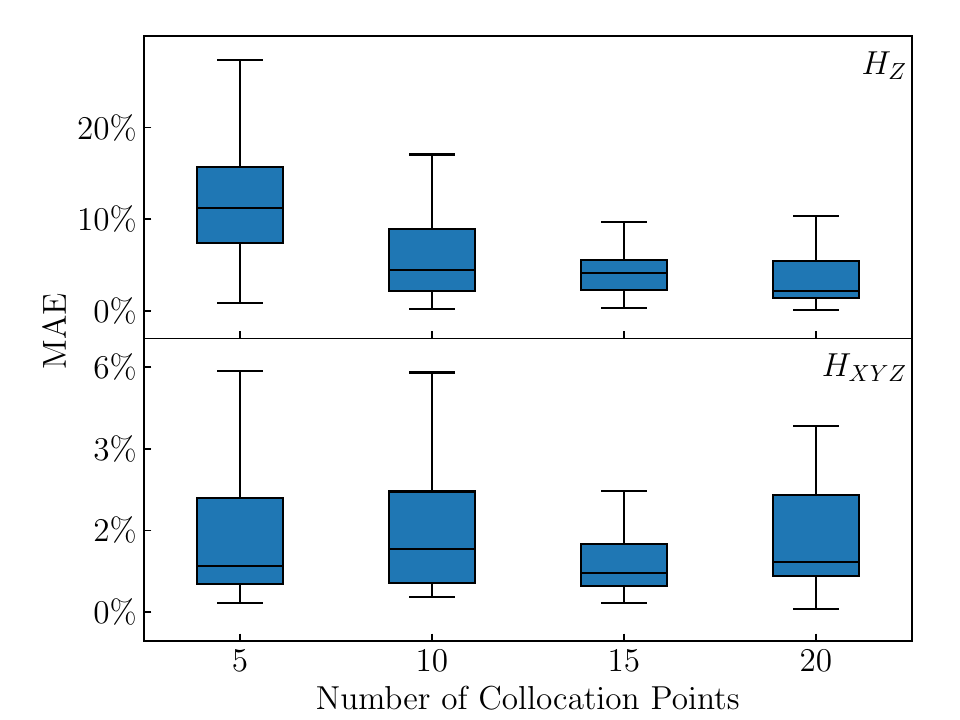}\label{fig4}
    \caption{MAE obtained from the experimental data provided by IBM quantum computers for the $H_Z$ Hamiltonian (top panel) and for the $H_{XYZ}$ Hamiltonian (bottom panel) as a function of the number of collocation points.}
    \end{figure}

In Figure \ref{fig2}, we plot the MAE as a function of the standard deviation $\sigma$ related to the Gaussian distribution of the errors for both Hamiltonians $H_Z$ (top panel) and $H_{XYZ}$ (bottom panel) considering only $N=5$ collocation points. In this case, we notice that the Hamiltonian tomography has at least 5\% accuracy if the standard deviation is less than 1\%. On the contrary, if the error has a 10\% standard deviation, the accuracy measured by MAE is very poor when using only $N=5$ collocation points. We repeated the same analysis by increasing the number of collocation numbers (results not shown here) and we found that it is possible to improve the Hamiltonian accuracy by increasing the number of collocations number, {\it e.g.}, we can get MAE $<7\%$  for $\sigma=10\%$ when $N=100$. After checking that the inverse-PINN provides a good accuracy for the tomography of $H_Z$ and $H_{XYZ}$, we repeat the same analysis for the general two qubit Hamiltonian (Eq.~\ref{eq:hamiltiniano_ML}), whose results are shown in Figure ~\ref{fig3}. In the top panel of Figure~\ref{fig3}, we plot MAE versus collocation points for the input data provided without errors. We can conclude that at least N=10 collocation points are necessary for an accurate prediction of the physical parameters of the general Hamiltonian (Eq.~\ref{eq:hamiltiniano_ML}). In the bottom panel of Figure ~\ref{fig3}, we consider N=20 collocation points and we include the random Gaussian error in the input data. In this case, we can see that the MAE is less than 3\% for a standard deviation of 1\%, which is very similar to the results found for $H_Z$ and $H_{XYZ}$ Hamiltonians. This result demonstrates that the inverse-PINN can indeed learn the general Hamiltonian with good accuracy considering at least N=20 collocation points, even though errors in the measurements are present. As a final test, we simulate the $H_Z$ the $H_{XYZ}$ Hamiltonians in the IBM quantum computers. We use these two Hamiltonians because of their easy representation in terms of quantum gates, which do not require any Trotter approximation. In Figure ~\ref{fig4}, we plot the MAE for both $H_Z$ and $H_{XYZ}$ Hamiltonians as a function of the number of collocation points. The difference of these results from the previous ones is related to the input data, which was experimentally obtained from the IBM quantum computers. For the $H_Z$ Hamiltonian (top panel of Figure ~\ref{fig4}), we find that the MAE only achieves values less than 5\% for 20 collocation points. On the other hand, the results in the bottom panel of Figure ~\ref{fig4} for the $H_{XYZ}$ Hamiltonian, show that the MAE is already less than 5\% on average for only 5 collocation points. We believe that this difference might be due to errors introduced in the realization of the quantum gates in the IBM quantum computer.

\section{Conclusion}
We applied a technique called inverse-PINN, which is based on neural networks, to extract the parameters of a two-qubit Hamiltonian. First, we tested two Hamiltonians of interest, one containing only terms in the z-direction and the so called XYZ model without local terms. We found that the inverse-PINN is able to find the parameters of both Hamiltonians with great accuracy, when the input data is provided without errors. When we include errors, by considering a Gaussian dispersion in the parameters that generate the input data, we still are able to estimate the real parameters with 5\% accuracy, if we use at least 20 collocation points. We also simulate a full two-qubit Hamiltonian, where there are 15 parameters to be learned and the results for the general case are similar to those found for the particular Hamiltonians. Furthermore, we use the input data obtained from the IBM quantum computers for the $H_Z$ and $H_{XYZ}$ Hamiltonians as an real scenario for our approach. In this case, we also found results with very good accuracy, similar to those found using the results with Gaussian dispersion. We believe that this approach is very powerful and can be used in other important tasks related to the improvement of quantum computation platforms.

\begin{acknowledgements}
The authors are grateful for financial support by  the Brazilian Agencies FAPESP, CNPq and CAPES. 
LKC thanks to the Brazilian Agencies FAPESP (grant 2019/09624-3) for supporting this research. 
\end{acknowledgements}


\begin{thebibliography}{25}
	\expandafter\ifx\csname natexlab\endcsname\relax\def\natexlab#1{#1}\fi
	\expandafter\ifx\csname bibnamefont\endcsname\relax
	\def\bibnamefont#1{#1}\fi
	\expandafter\ifx\csname bibfnamefont\endcsname\relax
	\def\bibfnamefont#1{#1}\fi
	\expandafter\ifx\csname citenamefont\endcsname\relax
	\def\citenamefont#1{#1}\fi
	\expandafter\ifx\csname url\endcsname\relax
	\def\url#1{\texttt{#1}}\fi
	\expandafter\ifx\csname urlprefix\endcsname\relax\def\urlprefix{URL }\fi
	\providecommand{\bibinfo}[2]{#2}
	\providecommand{\eprint}[2][]{\url{#2}}
	
	\bibitem[{\citenamefont{Alipanahi et~al.}(2015)\citenamefont{Alipanahi, Delong,
			Weirauch, and Frey}}]{Alipanahi2015}
	\bibinfo{author}{\bibfnamefont{B.}~\bibnamefont{Alipanahi}},
	\bibinfo{author}{\bibfnamefont{A.}~\bibnamefont{Delong}},
	\bibinfo{author}{\bibfnamefont{M.~T.} \bibnamefont{Weirauch}},
	\bibnamefont{and} \bibinfo{author}{\bibfnamefont{B.~J.} \bibnamefont{Frey}},
	\bibinfo{journal}{Nature Biotechnology} \textbf{\bibinfo{volume}{33}},
	\bibinfo{pages}{831} (\bibinfo{year}{2015}), ISSN \bibinfo{issn}{1546-1696},
	\urlprefix\url{https://doi.org/10.1038/nbt.3300}.
	
	\bibitem[{\citenamefont{LeCun et~al.}(2015)\citenamefont{LeCun, Bengio, and
			Hinton}}]{LeCun2015}
	\bibinfo{author}{\bibfnamefont{Y.}~\bibnamefont{LeCun}},
	\bibinfo{author}{\bibfnamefont{Y.}~\bibnamefont{Bengio}}, \bibnamefont{and}
	\bibinfo{author}{\bibfnamefont{G.}~\bibnamefont{Hinton}},
	\bibinfo{journal}{Nature} \textbf{\bibinfo{volume}{521}},
	\bibinfo{pages}{436} (\bibinfo{year}{2015}), ISSN \bibinfo{issn}{1476-4687},
	\urlprefix\url{https://doi.org/10.1038/nature14539}.
	
	\bibitem[{\citenamefont{Lake et~al.}(2015)\citenamefont{Lake, Salakhutdinov,
			and Tenenbaum}}]{doi:10.1126/science.aab3050}
	\bibinfo{author}{\bibfnamefont{B.~M.} \bibnamefont{Lake}},
	\bibinfo{author}{\bibfnamefont{R.}~\bibnamefont{Salakhutdinov}},
	\bibnamefont{and} \bibinfo{author}{\bibfnamefont{J.~B.}
		\bibnamefont{Tenenbaum}}, \bibinfo{journal}{Science}
	\textbf{\bibinfo{volume}{350}}, \bibinfo{pages}{1332} (\bibinfo{year}{2015}),
	\eprint{https://www.science.org/doi/pdf/10.1126/science.aab3050},
	\urlprefix\url{https://www.science.org/doi/abs/10.1126/science.aab3050}.
	
	\bibitem[{\citenamefont{Krizhevsky et~al.}(2012)\citenamefont{Krizhevsky,
			Sutskever, and Hinton}}]{NIPS2012_c399862d}
	\bibinfo{author}{\bibfnamefont{A.}~\bibnamefont{Krizhevsky}},
	\bibinfo{author}{\bibfnamefont{I.}~\bibnamefont{Sutskever}},
	\bibnamefont{and} \bibinfo{author}{\bibfnamefont{G.~E.}
		\bibnamefont{Hinton}}, in \emph{\bibinfo{booktitle}{Advances in Neural
			Information Processing Systems}}, edited by
	\bibinfo{editor}{\bibfnamefont{F.}~\bibnamefont{Pereira}},
	\bibinfo{editor}{\bibfnamefont{C.}~\bibnamefont{Burges}},
	\bibinfo{editor}{\bibfnamefont{L.}~\bibnamefont{Bottou}}, \bibnamefont{and}
	\bibinfo{editor}{\bibfnamefont{K.}~\bibnamefont{Weinberger}}
	(\bibinfo{publisher}{Curran Associates, Inc.}, \bibinfo{year}{2012}),
	vol.~\bibinfo{volume}{25},
	\urlprefix\url{https://proceedings.neurips.cc/paper_files/paper/2012/file/c399862d3b9d6b76c8436e924a68c45b-Paper.pdf}.
	
	\bibitem[{\citenamefont{Denby}(1988)}]{DENBY1988429}
	\bibinfo{author}{\bibfnamefont{B.}~\bibnamefont{Denby}},
	\bibinfo{journal}{Computer Physics Communications}
	\textbf{\bibinfo{volume}{49}}, \bibinfo{pages}{429} (\bibinfo{year}{1988}),
	ISSN \bibinfo{issn}{0010-4655},
	\urlprefix\url{https://www.sciencedirect.com/science/article/pii/0010465588900045}.
	
	\bibitem[{\citenamefont{Barate~et al. and
			Collaboration}(1999)}]{Barateetal.1999}
	\bibinfo{author}{\bibfnamefont{R.}~\bibnamefont{Barate~et al.}}
	\bibnamefont{and} \bibinfo{author}{\bibfnamefont{T.~A.}
		\bibnamefont{Collaboration}}, \bibinfo{journal}{The European Physical Journal
		C - Particles and Fields} \textbf{\bibinfo{volume}{6}}, \bibinfo{pages}{555}
	(\bibinfo{year}{1999}), ISSN \bibinfo{issn}{1434-6052},
	\urlprefix\url{https://doi.org/10.1007/s100529801031}.
	
	\bibitem[{\citenamefont{Radovic et~al.}(2018)\citenamefont{Radovic, Williams,
			Rousseau, Kagan, Bonacorsi, Himmel, Aurisano, Terao, and
			Wongjirad}}]{Radovic2018}
	\bibinfo{author}{\bibfnamefont{A.}~\bibnamefont{Radovic}},
	\bibinfo{author}{\bibfnamefont{M.}~\bibnamefont{Williams}},
	\bibinfo{author}{\bibfnamefont{D.}~\bibnamefont{Rousseau}},
	\bibinfo{author}{\bibfnamefont{M.}~\bibnamefont{Kagan}},
	\bibinfo{author}{\bibfnamefont{D.}~\bibnamefont{Bonacorsi}},
	\bibinfo{author}{\bibfnamefont{A.}~\bibnamefont{Himmel}},
	\bibinfo{author}{\bibfnamefont{A.}~\bibnamefont{Aurisano}},
	\bibinfo{author}{\bibfnamefont{K.}~\bibnamefont{Terao}}, \bibnamefont{and}
	\bibinfo{author}{\bibfnamefont{T.}~\bibnamefont{Wongjirad}},
	\bibinfo{journal}{Nature} \textbf{\bibinfo{volume}{560}}, \bibinfo{pages}{41}
	(\bibinfo{year}{2018}), ISSN \bibinfo{issn}{1476-4687},
	\urlprefix\url{https://doi.org/10.1038/s41586-018-0361-2}.
	
	\bibitem[{\citenamefont{Bedolla et~al.}(2020)\citenamefont{Bedolla, Padierna,
			and Castañeda-Priego}}]{Bedolla_2021}
	\bibinfo{author}{\bibfnamefont{E.}~\bibnamefont{Bedolla}},
	\bibinfo{author}{\bibfnamefont{L.~C.} \bibnamefont{Padierna}},
	\bibnamefont{and}
	\bibinfo{author}{\bibfnamefont{R.}~\bibnamefont{Castañeda-Priego}},
	\bibinfo{journal}{Journal of Physics: Condensed Matter}
	\textbf{\bibinfo{volume}{33}}, \bibinfo{pages}{053001}
	(\bibinfo{year}{2020}),
	\urlprefix\url{https://dx.doi.org/10.1088/1361-648X/abb895}.
	
	\bibitem[{\citenamefont{Carrasquilla}(2020)}]{Juan_Carrasquilla}
	\bibinfo{author}{\bibfnamefont{J.}~\bibnamefont{Carrasquilla}},
	\bibinfo{journal}{Advances in Physics} \textbf{\bibinfo{volume}{5}},
	\bibinfo{pages}{1} (\bibinfo{year}{2020}).
	
	\bibitem[{\citenamefont{Hashimoto et~al.}(2018)\citenamefont{Hashimoto,
			Sugishita, Tanaka, and Tomiya}}]{PhysRevD.98.046019}
	\bibinfo{author}{\bibfnamefont{K.}~\bibnamefont{Hashimoto}},
	\bibinfo{author}{\bibfnamefont{S.}~\bibnamefont{Sugishita}},
	\bibinfo{author}{\bibfnamefont{A.}~\bibnamefont{Tanaka}}, \bibnamefont{and}
	\bibinfo{author}{\bibfnamefont{A.}~\bibnamefont{Tomiya}},
	\bibinfo{journal}{Phys. Rev. D} \textbf{\bibinfo{volume}{98}},
	\bibinfo{pages}{046019} (\bibinfo{year}{2018}),
	\urlprefix\url{https://link.aps.org/doi/10.1103/PhysRevD.98.046019}.
	
	\bibitem[{\citenamefont{Canabarro et~al.}(2019)\citenamefont{Canabarro,
			Fanchini, Malvezzi, Pereira, and Chaves}}]{PhysRevB.100.045129}
	\bibinfo{author}{\bibfnamefont{A.}~\bibnamefont{Canabarro}},
	\bibinfo{author}{\bibfnamefont{F.~F.} \bibnamefont{Fanchini}},
	\bibinfo{author}{\bibfnamefont{A.~L.} \bibnamefont{Malvezzi}},
	\bibinfo{author}{\bibfnamefont{R.}~\bibnamefont{Pereira}}, \bibnamefont{and}
	\bibinfo{author}{\bibfnamefont{R.}~\bibnamefont{Chaves}},
	\bibinfo{journal}{Phys. Rev. B} \textbf{\bibinfo{volume}{100}},
	\bibinfo{pages}{045129} (\bibinfo{year}{2019}),
	\urlprefix\url{https://link.aps.org/doi/10.1103/PhysRevB.100.045129}.
	
	\bibitem[{\citenamefont{Raissi et~al.}(2019)\citenamefont{Raissi, Perdikaris,
			and Karniadakis}}]{RAISSI2019686}
	\bibinfo{author}{\bibfnamefont{M.}~\bibnamefont{Raissi}},
	\bibinfo{author}{\bibfnamefont{P.}~\bibnamefont{Perdikaris}},
	\bibnamefont{and}
	\bibinfo{author}{\bibfnamefont{G.}~\bibnamefont{Karniadakis}},
	\bibinfo{journal}{Journal of Computational Physics}
	\textbf{\bibinfo{volume}{378}}, \bibinfo{pages}{686} (\bibinfo{year}{2019}),
	ISSN \bibinfo{issn}{0021-9991},
	\urlprefix\url{https://www.sciencedirect.com/science/article/pii/S0021999118307125}.
	
	\bibitem[{\citenamefont{Wang et~al.}(2021)\citenamefont{Wang, Wang, and
			Perdikaris}}]{doi:10.1126/sciadv.abi8605}
	\bibinfo{author}{\bibfnamefont{S.}~\bibnamefont{Wang}},
	\bibinfo{author}{\bibfnamefont{H.}~\bibnamefont{Wang}}, \bibnamefont{and}
	\bibinfo{author}{\bibfnamefont{P.}~\bibnamefont{Perdikaris}},
	\bibinfo{journal}{Science Advances} \textbf{\bibinfo{volume}{7}},
	\bibinfo{pages}{eabi8605} (\bibinfo{year}{2021}),
	\eprint{https://www.science.org/doi/pdf/10.1126/sciadv.abi8605},
	\urlprefix\url{https://www.science.org/doi/abs/10.1126/sciadv.abi8605}.
	
	\bibitem[{\citenamefont{Lu et~al.}(2021)\citenamefont{Lu, Meng, Mao, and
			Karniadakis}}]{doi:10.1137/19M1274067}
	\bibinfo{author}{\bibfnamefont{L.}~\bibnamefont{Lu}},
	\bibinfo{author}{\bibfnamefont{X.}~\bibnamefont{Meng}},
	\bibinfo{author}{\bibfnamefont{Z.}~\bibnamefont{Mao}}, \bibnamefont{and}
	\bibinfo{author}{\bibfnamefont{G.~E.} \bibnamefont{Karniadakis}},
	\bibinfo{journal}{SIAM Review} \textbf{\bibinfo{volume}{63}},
	\bibinfo{pages}{208} (\bibinfo{year}{2021}),
	\eprint{https://doi.org/10.1137/19M1274067},
	\urlprefix\url{https://doi.org/10.1137/19M1274067}.
	
	\bibitem[{\citenamefont{Chuang and
			Nielsen}(1997)}]{doi:10.1080/09500349708231894}
	\bibinfo{author}{\bibfnamefont{I.~L.} \bibnamefont{Chuang}} \bibnamefont{and}
	\bibinfo{author}{\bibfnamefont{M.~A.} \bibnamefont{Nielsen}},
	\bibinfo{journal}{Journal of Modern Optics} \textbf{\bibinfo{volume}{44}},
	\bibinfo{pages}{2455} (\bibinfo{year}{1997}),
	\urlprefix\url{https://www.tandfonline.com/doi/abs}.
	
	\bibitem[{\citenamefont{Altepeter et~al.}(2003)\citenamefont{Altepeter,
			Branning, Jeffrey, Wei, Kwiat, Thew, O'Brien, Nielsen, and
			White}}]{PhysRevLett.90.193601}
	\bibinfo{author}{\bibfnamefont{J.~B.} \bibnamefont{Altepeter}},
	\bibinfo{author}{\bibfnamefont{D.}~\bibnamefont{Branning}},
	\bibinfo{author}{\bibfnamefont{E.}~\bibnamefont{Jeffrey}},
	\bibinfo{author}{\bibfnamefont{T.~C.} \bibnamefont{Wei}},
	\bibinfo{author}{\bibfnamefont{P.~G.} \bibnamefont{Kwiat}},
	\bibinfo{author}{\bibfnamefont{R.~T.} \bibnamefont{Thew}},
	\bibinfo{author}{\bibfnamefont{J.~L.} \bibnamefont{O'Brien}},
	\bibinfo{author}{\bibfnamefont{M.~A.} \bibnamefont{Nielsen}},
	\bibnamefont{and} \bibinfo{author}{\bibfnamefont{A.~G.} \bibnamefont{White}},
	\bibinfo{journal}{Phys. Rev. Lett.} \textbf{\bibinfo{volume}{90}},
	\bibinfo{pages}{193601} (\bibinfo{year}{2003}),
	\urlprefix\url{https://link.aps.org/doi/10.1103/PhysRevLett.90.193601}.
	
	\bibitem[{\citenamefont{Merkel et~al.}(2013)\citenamefont{Merkel, Gambetta,
			Smolin, Poletto, C\'orcoles, Johnson, Ryan, and
			Steffen}}]{PhysRevA.87.062119}
	\bibinfo{author}{\bibfnamefont{S.~T.} \bibnamefont{Merkel}},
	\bibinfo{author}{\bibfnamefont{J.~M.} \bibnamefont{Gambetta}},
	\bibinfo{author}{\bibfnamefont{J.~A.} \bibnamefont{Smolin}},
	\bibinfo{author}{\bibfnamefont{S.}~\bibnamefont{Poletto}},
	\bibinfo{author}{\bibfnamefont{A.~D.} \bibnamefont{C\'orcoles}},
	\bibinfo{author}{\bibfnamefont{B.~R.} \bibnamefont{Johnson}},
	\bibinfo{author}{\bibfnamefont{C.~A.} \bibnamefont{Ryan}}, \bibnamefont{and}
	\bibinfo{author}{\bibfnamefont{M.}~\bibnamefont{Steffen}},
	\bibinfo{journal}{Phys. Rev. A} \textbf{\bibinfo{volume}{87}},
	\bibinfo{pages}{062119} (\bibinfo{year}{2013}),
	\urlprefix\url{https://link.aps.org/doi/10.1103/PhysRevA.87.062119}.
	
	\bibitem[{\citenamefont{Anshu et~al.}(2021)\citenamefont{Anshu, Arunachalam,
			Kuwahara, and Soleimanifar}}]{Anshu2021}
	\bibinfo{author}{\bibfnamefont{A.}~\bibnamefont{Anshu}},
	\bibinfo{author}{\bibfnamefont{S.}~\bibnamefont{Arunachalam}},
	\bibinfo{author}{\bibfnamefont{T.}~\bibnamefont{Kuwahara}}, \bibnamefont{and}
	\bibinfo{author}{\bibfnamefont{M.}~\bibnamefont{Soleimanifar}},
	\bibinfo{journal}{Nature Physics} \textbf{\bibinfo{volume}{17}},
	\bibinfo{pages}{931} (\bibinfo{year}{2021}), ISSN \bibinfo{issn}{1745-2481},
	\urlprefix\url{https://doi.org/10.1038/s41567-021-01232-0}.
	
	\bibitem[{\citenamefont{Bairey et~al.}(2019)\citenamefont{Bairey, Arad, and
			Lindner}}]{PhysRevLett.122.020504}
	\bibinfo{author}{\bibfnamefont{E.}~\bibnamefont{Bairey}},
	\bibinfo{author}{\bibfnamefont{I.}~\bibnamefont{Arad}}, \bibnamefont{and}
	\bibinfo{author}{\bibfnamefont{N.~H.} \bibnamefont{Lindner}},
	\bibinfo{journal}{Phys. Rev. Lett.} \textbf{\bibinfo{volume}{122}},
	\bibinfo{pages}{020504} (\bibinfo{year}{2019}),
	\urlprefix\url{https://link.aps.org/doi/10.1103/PhysRevLett.122.020504}.
	
	\bibitem[{\citenamefont{Li et~al.}(2020)\citenamefont{Li, Zou, and
			Hsieh}}]{PhysRevLett.124.160502}
	\bibinfo{author}{\bibfnamefont{Z.}~\bibnamefont{Li}},
	\bibinfo{author}{\bibfnamefont{L.}~\bibnamefont{Zou}}, \bibnamefont{and}
	\bibinfo{author}{\bibfnamefont{T.~H.} \bibnamefont{Hsieh}},
	\bibinfo{journal}{Phys. Rev. Lett.} \textbf{\bibinfo{volume}{124}},
	\bibinfo{pages}{160502} (\bibinfo{year}{2020}),
	\urlprefix\url{https://link.aps.org/doi/10.1103/PhysRevLett.124.160502}.
	
	\bibitem[{\citenamefont{Zubida et~al.}(2021)\citenamefont{Zubida, Yitzhaki,
			Lindner, and Bairey}}]{zubida2021optimal}
	\bibinfo{author}{\bibfnamefont{A.}~\bibnamefont{Zubida}},
	\bibinfo{author}{\bibfnamefont{E.}~\bibnamefont{Yitzhaki}},
	\bibinfo{author}{\bibfnamefont{N.~H.} \bibnamefont{Lindner}},
	\bibnamefont{and} \bibinfo{author}{\bibfnamefont{E.}~\bibnamefont{Bairey}},
	\emph{\bibinfo{title}{Optimal short-time measurements for hamiltonian
			learning}} (\bibinfo{year}{2021}), \eprint{2108.08824}.
	
	\bibitem[{\citenamefont{Hangleiter et~al.}(2021)\citenamefont{Hangleiter, Roth,
			Eisert, and Roushan}}]{hangleiter2021precise}
	\bibinfo{author}{\bibfnamefont{D.}~\bibnamefont{Hangleiter}},
	\bibinfo{author}{\bibfnamefont{I.}~\bibnamefont{Roth}},
	\bibinfo{author}{\bibfnamefont{J.}~\bibnamefont{Eisert}}, \bibnamefont{and}
	\bibinfo{author}{\bibfnamefont{P.}~\bibnamefont{Roushan}},
	\emph{\bibinfo{title}{Precise hamiltonian identification of a superconducting
			quantum processor}} (\bibinfo{year}{2021}), \eprint{2108.08319}.
	
	\bibitem[{\citenamefont{Yu et~al.}(2023)\citenamefont{Yu, Sun, Han, and
			Yuan}}]{Yu2023robustefficient}
	\bibinfo{author}{\bibfnamefont{W.}~\bibnamefont{Yu}},
	\bibinfo{author}{\bibfnamefont{J.}~\bibnamefont{Sun}},
	\bibinfo{author}{\bibfnamefont{Z.}~\bibnamefont{Han}}, \bibnamefont{and}
	\bibinfo{author}{\bibfnamefont{X.}~\bibnamefont{Yuan}},
	\bibinfo{journal}{{Quantum}} \textbf{\bibinfo{volume}{7}},
	\bibinfo{pages}{1045} (\bibinfo{year}{2023}), ISSN \bibinfo{issn}{2521-327X},
	\urlprefix\url{https://doi.org/10.22331/q-2023-06-29-1045}.
	
	\bibitem[{\citenamefont{Flammia and Wallman}(2020)}]{10.1145/3408039}
	\bibinfo{author}{\bibfnamefont{S.~T.} \bibnamefont{Flammia}} \bibnamefont{and}
	\bibinfo{author}{\bibfnamefont{J.~J.} \bibnamefont{Wallman}},
	\bibinfo{journal}{ACM Transactions on Quantum Computing}
	\textbf{\bibinfo{volume}{1}} (\bibinfo{year}{2020}),
	\urlprefix\url{https://doi-org.ez31.periodicos.capes.gov.br/10.1145/3408039}.
	
	\bibitem[{\citenamefont{Shulman et~al.}(2012)\citenamefont{Shulman, Dial,
			Harvey, Bluhm, Umansky, and Yacoby}}]{shulman2012demonstration}
	\bibinfo{author}{\bibfnamefont{M.~D.} \bibnamefont{Shulman}},
	\bibinfo{author}{\bibfnamefont{O.~E.} \bibnamefont{Dial}},
	\bibinfo{author}{\bibfnamefont{S.~P.} \bibnamefont{Harvey}},
	\bibinfo{author}{\bibfnamefont{H.}~\bibnamefont{Bluhm}},
	\bibinfo{author}{\bibfnamefont{V.}~\bibnamefont{Umansky}}, \bibnamefont{and}
	\bibinfo{author}{\bibfnamefont{A.}~\bibnamefont{Yacoby}},
	\bibinfo{journal}{science} \textbf{\bibinfo{volume}{336}},
	\bibinfo{pages}{202} (\bibinfo{year}{2012}).
	
\end{thebibliography}
\end{document}